\documentclass[12pt]{article}

\usepackage{epsfig,amsfonts,amssymb}
\newcommand{\beq}{\begin{equation}}
\newcommand{\eeq}{\end{equation}}
\newcommand{\beqa}{\begin{eqnarray}}
\newcommand{\eeqa}{\end{eqnarray}}
\newcommand{\beqar}{\begin{eqnarray*}}
\newcommand{\eeqar}{\end{eqnarray*}}

\newcommand{\ie}{{\it i.e.,}\ }
\newcommand{\reef}[1]{(\ref{#1})}

\begin{document}

\thispagestyle{empty}

\hfill{}

\hfill{}

\hfill{}

\hfill{hep-th/0603081}

\vspace{32pt}

\begin{center}
\textbf{\Large Black hole entropy as entanglement entropy:}\\
\textbf{\Large a holographic derivation}\\

\vspace{40pt}

Roberto Emparan

\vspace{12pt}
\textit{Instituci\'o Catalana de Recerca i Estudis Avan\c cats (ICREA)}\\
\textit{and}\\
\textit{Departament de F{\'\i}sica Fonamental}\\
\textit{Universitat de
Barcelona, Diagonal 647, E-08028, Barcelona, Spain}\\\vspace{6pt}
\texttt{emparan@ub.edu}
\end{center}

\vspace{40pt}

\begin{abstract} 

We study the possibility that black hole entropy be identified as
entropy of entanglement across the horizon of the vacuum of a quantum
field in the presence of the black hole. We argue that a recent proposal
for computing entanglement entropy using AdS/CFT holography implies that
black hole entropy can be exactly equated with entanglement entropy. The
implementation of entanglement entropy in this context solves all the
problems (such as cutoff dependence and the species problem) typically
associated with this identification. 

\end{abstract}

\setcounter{footnote}{0}

\newpage

\noindent\textbf{1. Introduction.} The vacuum of a quantum field in
equilibrium with a black hole contains correlations between points
inside and outside the horizon that are responsible for the thermal
behavior detected by outside observers \cite{unruh,israel}. It is then
plausible that black hole entropy receives a contribution, possibly
accounting for all of it, from the entropy of the density matrix of a
quantum field obtained by tracing over degrees of freedom beyond
the horizon, \ie the entropy of entanglement across the horizon
\cite{bkls} (see \cite{jac} for a discussion and references to the
subject). The leading contribution to the entropy of entanglement across
a geometric boundary does indeed come out naturally as proportional to
the area of the boundary \cite{bkls,sred}, and so it seems to reproduce
a main feature of the Bekenstein-Hawking formula\footnote{It will be
useful to keep $\hbar$ explicit in our equations.}
\beq\label{bhent}
S_{BH}=\frac{\mathcal{A}_{h}}{4G\hbar}\,. 
\eeq 
However, attempts to push this
identification further have typically faced a number of difficulties:

\begin{itemize} 

\item The entropy of entanglement is divergent due to correlations of
modes with arbitrarily short wavelengths close to the dividing boundary
\cite{bkls,thooft,sred}. If this is regularized with an ultraviolet
cutoff $\lambda_{UV}$ then the resulting entropy is strongly
cutoff-dependent, $S\sim \mathcal{A}/\lambda_{UV}^{D-2}$.

\item The species problem: if one wants to fully account for black hole
entropy as entanglement entropy, then the number of fields contributing
to the entanglement entropy appears to need a miraculous adjustment
(compounded by the cutoff dependence) in order to exactly match the
Bekenstein-Hawking formula \reef{bhent}.

\item The gravitational backreaction of the quantum fields is typically
neglected. This backreaction may not only distort the black hole area as
well as change the field's contribution to entanglement. It will also
generate higher-derivative corrections to the gravitational action that
will result in contributions to black hole entropy that are not
proportional to the area. 

\end{itemize}

While quantum contributions to black hole entropy of the form
$\mathcal{A}_h/\lambda_{UV}^{D-2}$ may be absorbed in a renormalization
of Newton's constant \cite{sussug}, it is much less clear whether all of
the black hole entropy can be understood in this way, through some
`induced gravity' mechanism \cite{jac,ff}.

Here we discuss how the application of a recently proposed prescription to
compute entanglement entropy \cite{ryta}, based on
AdS/CFT holography, implies that black hole entropy can be exactly
identified with the entropy of entanglement of quantum fields across the
black hole horizon. In the setup we devise, all of the above
difficulties are solved or circumvented in a simple manner.

The gist of our derivation is to have a black hole localized at the
boundary of AdS, and not at its center ---the latter leads to a
different notion of entanglement entropy in AdS/CFT \cite{malda,brus}
that we will discuss below, before the conclusions. The idea of having a
horizon at the boundary to study the entanglement of the CFT was
pioneered by \cite{hms}, who applied it to the problem of understanding
the entropy of deSitter space. By restricting themselves to a very
specific calculable case, namely deSitter in $1+1$ dimensions, the
authors of ref.~\cite{hms} were able to perform a field-theoretic
calculation of the entanglement entropy and compare it to the area
formula. Here we take a different route, and compute instead the
entanglement entropy using the proposal of \cite{ryta}. This method
allows us to treat much more general cases, and in particular to include
actual black holes in our arguments, but, as we discuss in the
concluding section, being a bulk-based calculation, it does not directly
clarify the microscopic origin of the entropy.

\bigskip

\noindent\textbf{2. Holographic calculation of entanglement entropy.} In
ref.~\cite{ryta} Ryu and Takayanagi propose, and give evidence for, the
following way to compute the entropy of entanglement for a (conformal)
quantum field theory that admits a gravitational dual. Say that the
$D$-dimensional field theory is defined on the conformal timelike
boundary $\partial M$ of a $(D+1)$-dimensional bulk spacetime $M$
(typically an AdS geometry), and that we wish to compute the entropy of
entanglement of a subsystem $A\subset \partial M$ bounded by a
$(D-2)$-dimensional spatial surface $\partial A$. Build then a
$(D-1)$-dimensional static minimal surface $\gamma_A$ in $M$ whose
boundary in $\partial M$ is $\partial A$. The entanglement entropy in
$A$ is then 
\beq
\label{holoent} S_A=\frac{\mathcal{A}(\gamma_A)}{4
G_{D+1} \hbar}\,, 
\eeq 
where $\mathcal{A}(\gamma_A)$ is the area of the
minimal surface $\gamma_A$ and $G_{D+1}$ is Newton's constant, both
defined in the bulk spacetime $M$. The divergences in the entanglement
entropy mentioned in the introduction are reflected here in the
generically infinite area of $\gamma_A$ as it extends to asymptotic
infinity. So far the proposal has been formulated only for static
surfaces, but an extension to stationary surfaces, as required when
rotation is present, might be possible. In this paper, though, we will
confine ourselves to static situations.

Although clearly inspired by the Bekenstein-Hawking formula
\reef{bhent}, this prescription is different from it in that
\reef{holoent} is in general intended to apply to spacelike surfaces
that are not horizons, in fact the boundary $\partial A$ is typically an
artificial division. However, given the similarity between \reef{bhent}
and \reef{holoent} it seems plausible that a connection between the two
should exist. Our aim is to describe a concrete set up where the
identification of black hole entropy with entanglement entropy, computed
via AdS/CFT, becomes manifest.

\bigskip

\noindent\textbf{3. Set up.} Ref.~\cite{ryta} applied the proposal
\reef{holoent} to cases where the bulk spacetime is essentially empty or
contains a black hole at its center (not at the boundary), which
corresponds to a thermal state of the quantum conformal field theory. We
want, instead, that the spacetime in which the field theory lives
contains a black hole, \ie we need a black hole at the boundary.

Another ingredient in our construction is the introduction of an
ultraviolet cutoff for the conformal field theory, which is needed in
order to render finite the entanglement entropy. This is naturally
incorporated in AdS/CFT by cutting off from the bulk geometry the
spacetime extending from a given timelike surface, ``the brane", out to
infinity. As a result of this, not only the $D$-dimensional CFT dual to
the bulk is cut off in the UV, but also bulk gravity now contains a
normalizable graviton mode localized on the brane \cite{rs2} that gives
rise, in the dual field theory, to $D$-dimensional gravity coupled to
the cutoff CFT. Thus, the classical $(D+1)$-dimensional bulk theory
containing such a brane provides a dual description of the dynamics on
the boundary brane satisfying the $D$-dimensional equations
\cite{gubser}
\beq\label{dualeqs}
G_{\mu\nu}=8\pi G_{D}\langle T_{\mu\nu}\rangle\,,
\eeq 
where the Einstein tensor $G_{\mu\nu}$ and the Newton constant $G_{D}$
are those induced on the boundary brane from the bulk, $G_{D}$ being
related to the bulk constant $G_{D+1}$ and AdS radius $R$ by
\beq\label{gg}
G_{D}=\frac{D-2}{R}\,G_{D+1}\,.
\eeq
$\langle T_{\mu\nu}\rangle$ is the expectation value of the renormalized
stress-energy tensor of the CFT, including all leading (planar)
corrections in a $1/N$ expansion. Hence the boundary theory accounts for
the exact gravitational backreaction of the quantum CFT at planar level.
The cutoff length of the CFT is equal to the AdS radius,
\beq\label{uvcutoff}
\lambda_{UV}=R\,. 
\eeq

This is the version of AdS/CFT duality that applies to the
Randall-Sundrum model with a single Planck brane (RS2). Exactly what is
the dual CFT depends on the specific construction at hand: we shall only
need to assume that it is possible to make sense of the duality at least
as long as bulk quantum corrections are negligible. The decomposition of
degrees of freedom in terms of a CFT coupled to $D$-dimensional gravity
is only sensible as an effective theory at energies well below the
cutoff. In this regime we do not need to deal with the specific nature
of the brane, which becomes relevant only if we need to know the
ultraviolet completion of the theory.

\begin{figure}
\begin{center}\leavevmode  %
\epsfxsize=7cm \epsfbox{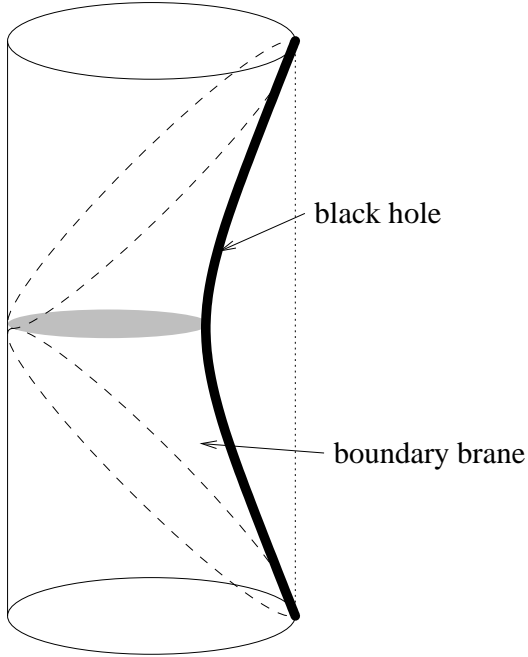}
\end{center}
\caption{\small Conformal diagram of the spacetime for a black hole on a
brane in AdS. On a spatial slice at constant time (shaded) the horizon
of the black hole is a minimal surface.
}
\label{fig:rs2bh}
\end{figure}

We shall now focus on $D=3$, since in this case there exists an explicit
exact bulk solution for a black hole localized on a RS2 three-brane in
AdS$_4$ \cite{ehm1} (see figure \ref{fig:rs2bh}). The metric induced on
the brane
\beq\label{3ds2}
ds^2|_\mathrm{brane}=-\left(1-\frac{r_0}{r}\right)dt^2+\frac{dr}{1-r_0/r}
+r^2d\phi^2\,
\eeq
contains a black hole horizon, as we
required. It can be seen that the angle $\phi$ has periodicity less than
$2\pi$ so there is a conical deficit that extends out to infinity
\cite{ehm1}. This geometry satisfies the three-dimensional Einstein
equations with stress-energy tensor
\beq\label{tmunu}
{T^\mu}_\nu=\frac{1}{16\pi G_{3}}\frac{r_0}{r^3}\,\mathrm{diag}(1,1,-2)\,.
\eeq
According to the discussion above, this stress-energy tensor must be
interpreted as the renormalized $\langle {T^\mu}_\nu\rangle$ of the CFT.

It is well known that there are no asymptotically flat black holes in
three-dimensional vacuum gravity, but instead point masses give rise to
locally flat conical spacetimes. Refs.~\cite{tanaka,efk} argued that, from the
point of view of the dual CFT+gravity 3D theory, the horizon in
\reef{3ds2} is the result of backreaction of the quantum conformal field
on a cone through eq.~\reef{dualeqs}. Ref.~\cite{efk} provided strong
evidence for this effect. In fact, if we write the horizon radius $r_0$
in terms of quantities of the dual 3D theory, including the
number of degrees of freedom $g_*$ of the CFT ,
\beq\label{gstar}
g_*\sim \frac{R}{\hbar G_3}=\frac{R^2}{\hbar G_4}\,,
\eeq
one finds $r_0=\hbar g_* G_3\, f(G_3M)$, indicating the quantum origin
of the horizon. Since we assume that we are in a regime where
$g_*$ is very large (in order to suppress bulk quantum gravity effects),
the horizon can have macroscopic size much larger than the 3D Planck
length $\hbar G_3$ or the CFT cutoff. 

Moreover, when expressed in terms of 3D quantities the entropy of the
black hole is proportional to $\hbar^0 g_*$. This strongly suggests
that the entropy of this black hole may have its origin in quantum
degrees of freedom of the CFT, possibly from entanglement of the vacuum
across the horizon. Since we do have
the explicit bulk dual of the black hole+CFT configuration
(\ref{3ds2},\ref{tmunu}), it seems natural to apply the formula
\reef{holoent} to the calculation of this entanglement entropy.

Before proceeding to this analysis, let us first identify the entropy of
the black hole \reef{3ds2}. The metric \reef{3ds2} is only the
brane-section of a four-dimensional bulk solution, given in \cite{ehm1},
with a black hole that extends into the bulk. Ref.~\cite{ehm1} argued
that the entropy of the black hole is given by the Bekenstein-Hawking
formula applied to the two-dimensional area $\mathcal{A}_{h}$ of the
black hole horizon {\em in the bulk}\footnote{This $S_{BH}$ differs from
\cite{ehm1} by a factor of 2, since we are taking the brane to be
single-sided instead of double-sided.},
\beq\label{sbrane}
S_{BH}=\frac{\mathcal{A}_{h}}{4G_{4}\hbar} 
=\frac{\pi R}{2G_3\hbar}\frac{x^2}{1+3x/2}\,,
\eeq
where $x$ is an auxiliary variable that satisfies $x^2(1+x)=r_0^2/R^2$,
and in the last equality we have expressed the result in terms of $G_3$
using \reef{gg}.
This is \textit{not} the same as the Bekenstein-Hawking entropy
associated to the one-dimensional length of the horizon circumference on
the brane,
$\mathcal{C}_{h}$,
\beq\label{sbh}
S_{\circ}=\frac{\mathcal{C}_{h}}{4G_3\hbar} 
= \frac{\pi R}{2G_3\hbar}\frac{x(x+1)}{1+3x/2}\,.
\eeq
The difference between \reef{sbrane} and \reef{sbh} is attributed to the
expected corrections to the effective action induced by renormalization
of the CFT. However, the two entropies become approximately the same in
the limit $r_0\gg R$ (\ie large $x$) in which the black hole is much
larger than the cutoff length of the CFT, in accordance with the
expectation that higher-derivative corrections to the action should be
suppressed in this limit. These corrections, conveniently expressed in
an expansion in $R/\mathcal{C}_{h}$, are
\beq\label{sbhso}
S_{BH}=S_{\circ}\left(1-\frac{4\pi}{3}\frac{R}{\mathcal{C}_{h}}
+\dots
\right)\,.
\eeq

\bigskip

\noindent\textbf{4. Black hole entropy equals entanglement entropy.}
Following the prescription \reef{holoent}, in order to compute the
entanglement entropy of the CFT in the region $A$ outside the black hole
horizon, we must find a minimal surface $\gamma_A$ in a spatial slice of
the bulk geometry whose boundary on the brane $\partial A$ coincides
with the horizon of the black hole on the brane. 

Since we are assuming staticity, this is actually very easy. On a
time-symmetric spatial slice, an apparent horizon (a surface where the
expansion of outgoing null geodesics vanishes) is a minimal
surface\footnote{More properly, an extremal surface. Under suitable
usual conditions it will also be minimal.}. So in a static spacetime, in
which the event horizon is also an apparent horizon, if we take a
spatial slice at constant Killing time the horizon will appear as a
minimal surface. Readers unfamiliar with this property of apparent
horizons in time-symmetric initial data sets may find this puzzling,
since it seems to go against the flat space intuition that a sphere at
constant radius is obviously not a minimal surface. That it is
nevertheless true for a black hole, can be easily visualized by
considering the section of the black hole on a spatial slice that goes
through the Einstein-Rosen bridge.

Therefore we identify the required $\gamma_A$ as the (spatial section at
constant Killing time of the) horizon in the bulk. It follows at once
that the entanglement entropy \reef{holoent} is the same as the entropy
of the black hole localized on the brane \reef{sbrane},
\beq\label{main}
S_{BH}=S_{A}(\partial A=\mathrm{horizon|_{brane}})\,
\eeq
(even if none of them is given by the area of the horizon on the
brane!).

The proof of this result is so simple that it immediately begs the
question of how it avoids the difficulties mentioned at the beginning of
the paper---concerns that this might be a trivial circular argument
will be addressed in the discussion at the end. 

The first two issues ---cutoff dependence and species problem--- are
basically dealt with together (similar points were made in \cite{hms}).
The RS2-AdS/CFT construction that we have used comes naturally equipped
with a UV cutoff for the CFT, \reef{uvcutoff}, that is at the same time
responsible for the strength of the gravitational coupling induced on
the brane $G_{3}$ through eq.~\reef{gg}. This same parameter fixes the
number of species of the dual CFT, in such a way that the effective
$G_{3}$ is correctly reproduced. In more detail, when working at length
scales much larger than the cutoff $R$, we expect the contribution of
each field to the entanglement entropy to be of the order of
\beq\label{onefield}
S_{(\mathrm{single\, field})}\sim \frac{\mathcal{C}_{h}}{R}\,,
\eeq
so for a number
$g_*$ of fields the total entanglement entropy is
\beq\label{}
S\sim g_*\frac{\mathcal{C}_{h}}{R}\sim
\frac{\mathcal{C}_{h}}{G_3\hbar}\,,
\eeq
where in the last line we have used \reef{gstar} (the numerical factors
come out correctly once we fix $g_*$ so that a thermal state of the CFT
reproduces the entropy of a black hole in AdS$_4$: this assumption
follows from the fact that we are always doing calculations in the same
side of the AdS/CFT duality). 
This adjustment of the number of degrees of freedom is similar to the
argument in the induced-gravity scenario, in which the effective Newton
constant $G_3$ also depends on the number of species \cite{jac}. 

Note that we have only included the entanglement of the CFT as the
primary contribution to the entropy. Quantum fluctuations of the
graviton on the brane (which result in quantum fuzzines in the position
of the horizon) will also be entangled across the horizon. However, in
the large $N$ (\ie large $g_*$) limit the number of modes of the
graviton is $O(1)$, compared to $g_*\gg 1$ for the CFT. The graviton
contribution is then subleading, and only becomes relevant when quantum
gravity in the bulk becomes important. Observe that in the regime in
which we work, the cutoff length $R$ of the CFT is much larger than
either the three- or four-dimensional Planck lengths, so the
entanglement entropy of each individual mode is much smaller than the
Bekenstein-Hawking entropy, $S_{(\mathrm{single\, field})}\ll
S_{\circ}\simeq S_{BH}$.

Finally, the gravitational backreaction of the CFT is incorporated
through the equation \reef{dualeqs} of RS2-AdS/CFT duality. The fact that the
surface $\gamma_A$ is the same as the spatial section of the bulk
horizon guarantees that the black hole entropy will be the same as the
entanglement entropy also after including corrections that change the
entropy away from the Bekenstein-Hawking entropy on the brane $S_{\circ}$.

\bigskip

\noindent\textbf{5. Comparison to entanglement on a circle in flat
space.} It is instructive to compare the entanglement across the black
hole horizon to the entanglement on a circle (since we are in two
spatial dimensions) of the same length in flat space. This is analogous
to the calculations in \cite{bkls,sred}, but now we use the holographic
method \reef{holoent} to compute the entanglement entropy.

\begin{figure}
\begin{center}\leavevmode  %
\epsfxsize=10cm \epsfbox{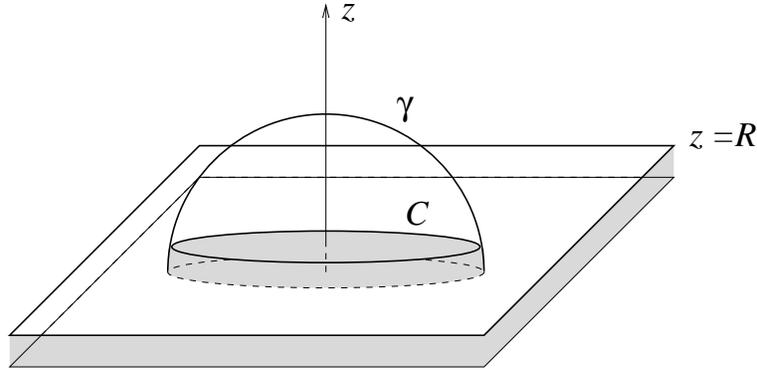}
\end{center}
\caption{\small Minimal surface $\gamma$ computing the entanglement
across a circle $C$ on a flat plane. A spatial section of AdS is
represented as the Poincare upper-half-volume.
The shaded region $0\leq z<R$ is cut off in the RS2 construction. The minimal
surface is a hemisphere, of which only the portion above $z=R$ is
relevant. The length of the circle $C$ is fixed to be $\mathcal{C}_{h}$.
The proper distance along $z$ is $R\log(z/R)$, so when
$R/\mathcal{C}_{h}\ll 1$ the hemisphere is actually pancaked along the
brane.
}
\label{fig:minsurf2}
\end{figure}

Therefore, we take AdS$_4$ with a RS2 two-brane that is flat (and without any
conical deficit), 
\beq
ds^2=\frac{R^2}{z^2}(-dt^2+dz^2+dr^2+r^2 d\phi^2)\,,\qquad z\geq R\,,
\eeq
and consider as our entanglement boundary a circle $C$ on the brane with
length equal to $\mathcal{C}_{h}$ (given by \reef{sbh}),
$r=\mathcal{C}_{h}/2\pi$. We must find the area of a minimal surface in
the bulk, $\gamma$, that is bounded by the circle $C$. This calculation
has actually been done in \cite{ryta}, and results in an entanglement
entropy equal to
\beqa\label{sc}
S_C&=&\frac{\mathcal{C}_{h}}{4G_3\hbar}\left(\sqrt{1+ \left(\frac{2\pi
R}{\mathcal{C}_{h}}\right)^2} -\frac{2\pi
R}{\mathcal{C}_{h}}\right)\nonumber\\
&=&S_{\circ}\left(1-\frac{2\pi R}{\mathcal{C}_{h}}
+\dots
\right)\,,
\eeqa
where we have expanded for small $R/\mathcal{C}_{h}$ to compare with
\reef{sbhso}. We see that the leading term agrees in both cases. The
reason that the agreement includes the precise factors, without any
ambiguities from cutoff dependence, is the same as explained in the
preceding section. Geometrically, it is easy to see (figure
\ref{fig:minsurf2}) that the minimal surface $\gamma$ is `pancaked'
along the brane in a manner very similar to that of horizons of large
black holes in RS2 \cite{ehm1}, so their total bulk area is $\simeq
\mathcal{C}_{h} R$ in both cases. The factor of $R$ (the cutoff length)
in this bulk area is then absorbed into $G_3$ in the computation of the
entropy.

The subleading corrections to the entropy in \reef{sc}, which are again
interpreted as due to higher-derivative terms in the action, differ by a
factor $2/3$ from the corrections in \reef{sbhso}. This numerical
discrepancy is not surprising, but it is interesting to see that the
sign of the corrections is the same in both cases. 

In the opposite limit of small black holes, $R/\mathcal{C}_{h}\gg 1$,
the entropy $S_C$ can only account for half of $S_{BH}$ (one half being
the ratio between the areas of a disk and a hemisphere of the same
radius), again not unexpected in a regime dominated by
higher-derivative operators. Note however that the identity \reef{main}
between $S_{BH}$ and $S_{A}$ on the horizon holds irrespective of the
size of the black hole.

\bigskip

\noindent\textbf{6. Generalization to higher dimensions.} No exact
solution for a static black hole with regular horizon on a RS2 brane in
AdS$_{D>4}$ is known. Refs.~\cite{tanaka,efk} argue that indeed no such
solution should exist, since the backreaction of the quantum fields
should make the black hole shrink as a consequence of Hawking
evaporation. However, the possibility that strong coupling effects of
the CFT complicate the picture seems to leave some room for skeptics
\cite{rob}. Within this paper we can remain agnostic as to the final
resolution of this problem, and discuss both possibilities. The
conclusion that entanglement entropy is the primary source for black
hole entropy is not essentially affected by this issue.

If static regular black holes on RS2 branes in AdS$_{D>4}$ exist,
it is clear that, since in any dimension the horizon is a minimal
surface on a slice at constant time, then the black hole entropy is in
general equal to the holographic definition of the entanglement entropy
of the CFT, \reef{main}. A holographic calculation of the entanglement
entropy on a sphere in flat space of the same area as the black hole
horizon, will again reproduce precisely the leading contribution
$S_{\circ}$ to the black hole entropy. 

If, on the contrary, black holes on the brane are necessarily
time-dependent due to Hawking emission of dual CFT radiation, then the
situation becomes more complicated since the outgoing radiation is also
expected to be correlated with the black hole state. Still, the time
evolution may be slow enough to allow for an approximate holographic
definition of entanglement entropy. In the regime of validity of the
effective CFT+gravity theory, in which the black hole is much larger
than the cutoff size, the evaporation rate should in fact be small. So
in this regime we also expect that entanglement entropy, possibly
computed from a minimal surface very close to the horizon at some
constant-time slice, accounts for most of the black hole entropy.
However, we will not attempt here to make any of these notions more
precise. It should be very interesting to study how the information
about correlations escapes out of the black hole. 

Finally, we mention that, if in addition to the ultraviolet cutoff, an
infrared cutoff brane is introduced, then it is possible to have regular
black holes in the bulk in the form of black strings extending between
the UV and IR branes (the distance between them must be closer than $R$
in order to avoid dynamical instabilities). In this case, our general
holographic argument applies again to prove that black hole entropy is
exactly equal to entanglement entropy.

\bigskip

\noindent\textbf{7. Connection to another notion of
entanglement within AdS/CFT}. Our approach to relate
black hole entropy to entanglement entropy using AdS/CFT holography
looks conceptually very different from another approach discussed
recently \cite{malda} (see also \cite{brus}). The notion of entanglement
considered in \cite{malda} involves a configuration where a black hole
is {\it at the center of AdS}, instead of at its boundary. The maximally
extended spacetime (the `eternal black hole') contains two causally
disconnected asymptotically AdS regions, 1 and 2, each with its own
causal boundary. So there are two copies of the dual CFT ---CFT$_1$ and
CFT$_2$---, and the total Hilbert space is ${\cal H}={\cal
H}_1\times{\cal H}_2$. This is a thermofield double: the total state of
the CFT is a pure state, but if we trace
over the degrees of freedom in ${\cal H}_2$ then we obtain a thermal
density matrix for the state in CFT$_1$. In contrast to the situation we
have been discussing, these two copies of the CFT do not share any
common geometric dividing surface: the boundary where the CFT is defined
does not contain a black hole.

It is however possible to relate this notion of entanglement to the one
we have studied here. To this effect, we follow \cite{hms} and imagine
that we gradually move the bulk black hole, initially at the center of
AdS, towards the boundary brane\footnote{This need not be an actual
evolution in time, but may instead be regarded as a one-parameter
sequence of eternal black hole spacetimes, in each of which the black
hole is moving about the center of AdS with increasing amplitude of
oscillation. In this way we can more easily work with the concept of two
causally disconnected asymptotic regions.}. As the bulk black hole
approaches the brane, the dual view from the boundary is that a
spherical `blob' of thermal conformal fields is contracting. The
maximally extended spacetime always contains a second asymptotic region, at
whose boundary a CFT$_2$ is defined. This is entangled with the CFT$_1$
so the total state of the field is always pure. 

When the black hole reaches the brane and sticks to it (possibly only
temporarily), the dual description is that the thermal conformal field
collapses and forms a black hole. This is the kind of black hole
localized on the brane that we have been studying in the previous
sections. Now, this black hole on the boundary brane can again be
maximally extended across the horizon to another asymptotic region on
the brane. It is in this second asymptotic region at the boundary where
we find the CFT$_2$. But now notice that the two CFTs live in the
presence of a black hole, one at each side of the horizon. The total
state of the CFT is a pure, Hartle-Hawking vacuum, and tracing over
CFT$_2$ we obtain a thermal state for the CFT$_1$ that lives in `our'
side of the horizon. This is exactly the same thermofield double in the
presence of a black hole spacetime as discussed in \cite{israel}.

Roughly speaking, as the black hole moves towards the brane, the CFT$_2$
is `carried along' with it, so the CFT$_1$ and CFT$_2$ are gradually
getting together. When the black hole reaches the brane, the two
boundaries where the CFTs are defined become the two asymptotic regions
of the brane section of the black hole.

In this manner we interpolate between the pictures in \cite{malda} and
in this paper, of two entangled copies of the CFT. At all times the
entropy of the black hole arises from this entanglement, the only
difference being in whether the boundary where the CFT is defined
contains the black hole horizon or not.

\bigskip

\noindent\textbf{8. Discussion.} The extreme simplicity of the argument
for the main result in this paper, eq.~\reef{main}, may prompt the
criticism that the reasoning might actually be circular, assuming already what
it purports to prove ---an entropy equal to an area divided by $4G$.
This is not the case. The holographic definition \reef{holoent} is
certainly inspired by the Bekenstein-Hawking formula, but it is actually
different from it and is designed to compute a magnitude that is
independent of the possible presence of a black hole (either in the bulk
or the boundary) and which has a definite interpretation in the dual CFT
theory even in the absence of gravity. To repeat it simply and precisely,
our claim, not more nor less, is that the proposal of \cite{ryta} for
calculating the entanglement entropy of a field theory using its
gravitational dual, \reef{holoent}, does imply that black hole entropy
can be interpreted as entropy of entanglement.

The main strength of our derivation is its generality: it applies to any
$D$-dimensional system of a black hole, and quantum fields in its
presence, that admits a $(D+1)$-dimensional gravitational dual.
Admittedly, the directness of the argument leading to \reef{main} stems
from the fact that all our calculations are done in the same side of the
AdS/CFT duality---namely, the AdS side. This does not trivialize the
conclusion, but it limits the insights we can obtain. In particular, we
do not claim to have provided a microscopic counting of black hole
entropy as entanglement entropy, although our work definitely supports
the idea that such a calculation should in principle be possible. A
microscopic counting would consist in the computation, within quantum
field theory, of the entanglement entropy of the CFT across the horizon
of a black hole that solves the equations \reef{dualeqs}. As such, this
would be a test of the prescription \reef{holoent}. For deSitter space
in $D=2$, where the dual bulk solution is very simple and the
divergences of the entanglement entropy are only logarithmic, this
calculation was carried out successfully in \cite{hms}. It seems very
difficult, though, to extend these results beyond two-dimensional
deSitter. For the case $D=3$ analyzed in most detail here, besides the
problems in controlling the divergences, our lack of knowledge about the
theory on a stack of M2-branes dual to AdS$_4 (\times S^7)$, and more
generally, the difficulties associated to field theory calculations at
strong coupling, make a microscopic calculation including precise
numerical factors unfeasible in practice. 

Nevertheless, the AdS/CFT duality has been subject to extensive tests,
and the formula \reef{holoent} seems to hold its ground very well too,
so one may be tempted to assume it and conclude from our results that it
is indeed possible to view black hole entropy as entanglement entropy.
At the very least, we think we have provided new grounds in support of
this interpretation.

\bigskip


\noindent\textbf{Acknowledgements.} I would like to thank Jaume Garriga
and Tony Padilla for a helpful discussion. I also thank Ted Jacobson for
pointing me to ref.~\cite{hms}. This work is partially supported by
grants CICyT FPA2004-04582-C02-02, European Comission FP6 program
MRTN-CT-2004-005104, and DURSI 2005SGR 00082.

\end{document}